\documentclass[aip,graphicx]{revtex4-1}

\usepackage[dvipdfmx]{graphicx}
\usepackage{color}
\usepackage{amsmath}
\usepackage{dcolumn}
\usepackage{bm}
\usepackage{multirow}

\draft % marks overfull lines with a black rule on the right

\begin{document}
\title{Field-tunable Weyl points and large anomalous Hall effects in degenerate magnetic semiconductor EuMg$_2$Bi$_2$}

\author{M. Kondo}\email[Corresponding author: ]{kondo@gmr.phys.sci.osaka-u.ac.jp}
\affiliation{Department of Physics, Osaka University, Toyonaka, Osaka 560-0043, Japan}
\author{M. Ochi}
\affiliation{Department of Physics, Osaka University, Toyonaka, Osaka 560-0043, Japan}
\affiliation{Forefront Research Center, Osaka University, Toyonaka, Osaka 560-0043, Japan}
\author{R. Kurihara}
\affiliation{The Institute for Solid State Physics, The University of Tokyo, Kashiwa, Chiba 277-8581, Japan}
\affiliation{Department of Physics, Faculty of Science and Technology, Tokyo University of Science, Noda, Chiba 278-8510, Japan}
\author{A. Miyake}
\affiliation{The Institute for Solid State Physics, The University of Tokyo, Kashiwa, Chiba 277-8581, Japan}
\author{Y. Yamasaki}
\affiliation{Research and Services Division of Materials Data and Integrated System (MaDIS), National Institute for Materials Science(NIMS), Tsukuba, Ibaraki 305-0047, Japan}
\affiliation{Center for Emergent Matter Science (CEMS), RIKEN, Wako, Saitama 351-0198, Japan}
%\affiliation{PRESTO, Japan Science and Technology Agency, Kawaguchi, Saitama 332-0012, Japan}
%
\author{M. Tokunaga}
\affiliation{The Institute for Solid State Physics, The University of Tokyo, Kashiwa, Chiba 277-8581, Japan}
\author{H. Nakao}
\affiliation{Condensed Matter Research Center and Photon Factory, Institute of Materials Structure Science, KEK, Tsukuba, Ibaraki 305-0801, Japan}
\author{K. Kuroki}
\affiliation{Department of Physics, Osaka University, Toyonaka, Osaka 560-0043, Japan}
\author{T. Kida}
\affiliation{Center for Advanced High Magnetic Field Science (AHMF), Graduate School of Science, Osaka University, Toyonaka, Osaka 560-0043, Japan}
\author{M. Hagiwara}
\affiliation{Center for Advanced High Magnetic Field Science (AHMF), Graduate School of Science, Osaka University, Toyonaka, Osaka 560-0043, Japan}
\author{H. Murakawa}
\affiliation{Department of Physics, Osaka University, Toyonaka, Osaka 560-0043, Japan}
\author{N. Hanasaki}
\affiliation{Department of Physics, Osaka University, Toyonaka, Osaka 560-0043, Japan}
\author{H. Sakai}\email[Corresponding author: ]{sakai@phys.sci.osaka-u.ac.jp}
\affiliation{Department of Physics, Osaka University, Toyonaka, Osaka 560-0043, Japan}

\date{\today}

\begin{abstract}
{
Magnets, with topologically-nontrivial Dirac/Weyl points, have recently attracted significant attention owing to the unconventional physical properties, such as large anomalous Hall effects. 
However, they typically have a high carrier density and complicated band structure near the Fermi energy. 
In this study, we report degenerate magnetic semiconductor EuMg$_2$Bi$_2$, which exhibits a single valley at the $\Gamma$ point, where the field-tunable Weyl points form via the magnetic exchange interaction with the local Eu spins. 
By the high-field measurements on high-quality single crystals, we observed the quantum oscillations in resistivity, elastic constant, and surface impedance, which enabled us to determine the position of the Fermi energy. 
In combination with the first-principles calculation, we revealed that the Weyl points are located in the vicinity of the Fermi energy when the Eu spins are fully polarized. 
Furthermore, we observed large anomalous Hall effect (Hall angle $\Theta_{\rm AH} \sim$0.07) in the forced ferromagnetic phase, which is consistent with this field variation of band structure.
%where the Hall angle is $\sim$0.07.
}
\end{abstract}

\maketitle

\section{INTRODUCTION}
The interplay of magnetism and topology is one of the most active research topics in recent condensed matter physics.
A magnetic Weyl semimetal or Weyl magnet is a typical topological magnet, in which the spin-polarized linear bands cross in the vicinity of the Fermi energy, and thereby, forming topologically protected crossing points (Weyl points).
Since large Berry curvature originating from the Weyl points leads to exotic physical properties \cite{MWSM_review} such as giant anomalous Hall/Nernst effects (AHE/ANE) and chiral anomalies, exploration of Weyl magnets is important not only for basic science but also for device applications.
To date, a number of Weyl magnets have been predicted and their experimental verifications are currently in progress.
Typical examples include antiferromagnets with non-collinear magnetic structures (e.g., Mn$_3$(Sn,Ge)\cite{Mn3Sn_AHE_Nakatsuji, Mn3Sn_ANE_Nakatsuji, Mn3Ge_AHE_Nakatsuji, Mn3X, Nayak2016SA, Xu2020SA}) and ferromagnets with peculiar crystal structures such as Kagome and Heusler lattices (e.g., Co$_3$Sn$_2$S$_2$,\cite{Co3Sn2S2_AHE_Felser, Co3Sn2S2_Wang_AHE, Co3Sn2S2_AHE_TokuraGp, Co3Sn2S2_ANE} Fe$_3$GeTe$_2$\cite{Fe3GeTe2_AHE_Kim, Fe3GeTe2_ANE}, Co$_2$Mn(Al,Ga)\cite{Co2MnGa_ANE_Nakatsuji, Co2MnGa_ANE_Felser, Co2MnGa_AHE, Xu2020PRBR, Sumida2020ComMater}).
These metallic magnets generally exhibit complicated band structures with many topologically-trivial bands near the Fermi energy.

%%%%%
\par
%%%%%

Another way to obtain a Weyl magnet involves using magnetic interaction in semimetals or narrow-gap semiconductors with partial substitution of magnetic elements (e.g., Eu$^{2+}$, Gd$^{3+}$) acting as localized spins.
In this case, given that Weyl points are formed by the exchange splitting of the bands due to the (field-induced) magnetization, materials with much simpler band structures can be considered as the candidates.
As typical examples, in GdPtBi\cite{GPB_Felser, GPB_Suzuki}, EuTiO$_3$\cite{KSTakahashi2019SA}, EuCd$_2$(As, Sb)$_2$\cite{EuCd2As2_AHE_PRB, EuCd2As2_AHE_arXiv, EuCd2As2_PRB, EuCd2Sb2_Uchida, EuCd2Sb2_APL} and $\alpha$-EuP$_3$\cite{EuP3_Mayo}, various unconventional AHE associated with the field-induced Weyl points have been recently reported.
However, the number of such systems is still limited and further development is required.

%%%%%
\par
%%%%%

We here focused on the $X$Mg$_2$Bi$_2$ compound ($X=$Ca, Sr, Ba, Eu and Yb)\cite{XMB_structure, BaMg2Bi2_ARPES_Sato, EuMg2Bi2_ARPES_JAP, XMB_calc_HSE06, XMB_band_calc, EMgB_neutron, EuMg2Bi2_Pakhira}.
This compound exhibits the CaAl$_2$Si$_2$-type crystal structure [space group $P\bar{3}m1$, see Fig.1(e) inset], which corresponds to alternative stacking of the $X$ layer with a triangular lattice and Mg$_2$Bi$_2$ layer with a buckled honeycomb lattice.
The previous first-principles calculation predicted that the valence band top and conduction band bottom, mainly consisting of Mg $s$ and Bi $p$ orbitals, are located very close to the Fermi energy at the $\Gamma$ point and no other bands exist near the Fermi energy\cite{XMB_band_calc, XMB_calc_HSE06}.
Furthermore, it was highlighted that the band gap (and possible band inversion) is tunable with the $X$ species.
Recent angel-resolved photoemission spectroscopy (ARPES) experiments indeed revealed signature of a transition from a band insulator to a Dirac semimetal via replacement of the $X$ site from Sr to Ba\cite{BaMg2Bi2_ARPES_Sato}.
Hence, this type of band structure, which is sensitive to the $X$ sites, should also be significantly modified via the exchange interaction, when the non-magnetic $X$ site is substituted with magnetic Eu$^{2+}$ ion.
It was previously reported that EuMg$_2$Bi$_2$ undergoes antiferromagnetic order below $\sim$6.7 K\ at zero magnetic field\cite{EMgB_neutron, EuMg2Bi2_Pakhira}.
Moreover, the band structure at zero magnetic field was examined via the theoretical calculation and ARPES, which revealed a single small hole pocket around the $\Gamma$ point\cite{EuMg2Bi2_ARPES_JAP}.
However, its variation with respect to the magnetic states of Eu spins and associated magnetotransport phenomena have been elusive.
\par
In this study, by the detailed transport measurement and first-principles calculation, we show the band structure dependent on the magnetic order for EuMg$_2$Bi$_2$.
Specifically, we observed clear quantum oscillation originating from the bulk band at high fields (up to 55 T), which enabled us to experimentally determine the Fermi energy by comparing with the calculated band structure.
Hence, it was clarified that the Weyl points are located in the vicinity of the Fermi energy, when Eu spins are fully polarized.
In experiments, we observed large AHE in the forced ferromagnetic phase, which was comparable in magnitude to those reported for ferromagnetic Weyl (semi)metals.
This suggests significant impacts of the emergent Weyl points.

\section{EXPERIMENTAL and computational METHODS}
Single crystals of EuMg$_2$Bi$_2$ were grown via a self-flux method\cite{XMB_structure} with starting composition of EuMg$_8$Bi$_{10}$.
Large single crystals with a typical size corresponding to 5 $\times$ 5 $\times$ 5 mm$^3$ were obtained [see inset of Fig.\,1(h)].
We confirmed that EuMg$_2$Bi$_2$ exhibits the tetragonal CaAl$_2$Si$_2$-type structure with space group $P\bar{3}m1$ [$a=b=4.7781(3)$\AA, $c=7.8557(6)$\AA] via the powder X-ray diffraction performed using Rigaku MiniFlex 600 X-ray diffractometer with Cu $K\alpha$ radiation ($\lambda=0.15418$ nm) at 300 K (Fig. S1 in Supplementary Materials).
The resonant X-ray magnetic diffraction measurements near the Eu $L_3$ absorption edge ($E=6.975$ keV) were performed at BL-3A in Photon Factory, KEK, Japan.
%%%%%
\par
%%%%%
The low-field transport properties were measured by a conventional 4-probe AC method using a Physical Property Measurement System (PPMS, Quantum Design).
The field dependences of magnetization were measured using a PPMS with AC measurement system option.
We used a Magnetic Property Measurement System (MPMS, Quantum Design) with Reciprocating Sample Option (RSO) for the temperature dependences of the magnetic susceptibility.
%%%%%
\par
%%%%%
The high-field measurements were performed up to 55 T using the non-destructive mid-pulse magnet with a 900 kJ capacitor bank at the International MegaGauss Science Laboratory at the Institute for Solid State Physics, University of Tokyo.
For resistivity measurements, we used a 4-probe AC method with a typical current and frequency of 5 mA and 50 kHz, respectively.
The typical sample size used in the resistivity measurement was 1$\times$0.3$\times$0.1 mm$^3$.
Furthermore, we performed the quantum oscillations measurements using a tunnel diode oscillator (TDO)\cite{TDO_methods} and an ultrasonic pulse-echo method.
For the former measurements, the sample was placed in one of two circles of a 0.8 mm-diameter 8-shaped coil to cancel the induced voltage due to the field change. 
The coil is a part of a TDO circuit, which resonates at $\sim$103 MHz at 4.2 K.
This resonant frequency is reduced to $\sim$1 MHz using a superheterodyne circuit.
The frequency $\Delta f$ measures the surface impedance of the metallic sample.
Thus, the oscillation in $\Delta f$ reflects the Shubnikov-de Haas (SdH) oscillation.
For latter measurements, an ultrasonic pulse-echo method, with a numerical vector-type phase detection technique, \cite{ultraasonic_method} was used to measure the longitudinal ultrasonic velocity $v_{33}$. 
The elastic constant, $C_{33}$ = $\rho v_{33}^2$, was determined from $v_{33}$ and mass density $\rho$. 
Piezoelectric transducers using LiNbO$_3$ plates, with a 36$^{\circ}$ Y-cut and fundamental frequency of approximately $f = 30$ MHz were employed to generate the higher-harmonic frequency of 104 MHz. 
A room-temperature vulcanizing rubber (Shin-Etsu Silicone KE-42T) was used to glue the LiNbO$_3$ onto the sample.
%%%%%
\par
%%%%%
We performed first-principles band structure calculations based on the density functional theory with the HSE06 energy functional~\cite{HSE06} and projector augmented wave methods~\cite{paw} as implemented in the Vienna {\it ab initio} simulation package~\cite{vasp1,vasp2,vasp3,vasp4}.
We used the experimental crystal structure determined at room temperature in our experiment.
To represent the magnetic structure, we used the unit cell doubled along the $c$-axis.
We also applied the $+U$ correction~\cite{DFTU1} with the simplified rotationally invariant approach~\cite{DFTU2} to the Eu-$f$ orbitals with $U_{eff}\equiv U-J = 5$ eV. 
%For the calculation in the canted antiferromagnetic phase, the magnetic moments were set in the $bc$-plane, and the resultant cant angle corresponded to 30.0$^{\circ}$ with respect to the $b$ axis.
The core electrons in the PAW potential are [Kr]4$d^{10}$ for Eu, [He] for Mg, and [Xe]$4f^{14}5d^{10}$ for Bi.
The plane-wave cutoff energy of 500 eV and an $8\times 8\times 4$ $\bm{k}$-mesh were used with the inclusion of the spin-orbit coupling.
After the first-principles calculation, we constructed the Wannier functions~\cite{Wannier1,Wannier2} using the \textsc{Wannier90} software~\cite{Wannier90}.
To prevent the mixture of the different spin components, we did not perform the maximal localization procedure.
We employed the Bi-$p$ orbitals as the Wannier basis. 
An $8\times 8\times 4$ $\bm{k}$-mesh was used for the Wannier construction.
By using the tight-binding model consisting of the aforementioned Wannier functions, we obtained the band structure with the spin polarization, $\langle S_z \rangle$.
To calculate the cross section of Fermi surface and the carrier density using our tight-binding model, we employed a $250\times 250$ $\bm{k}$-mesh on the $k_z=0$ plane and a $300\times 300\times 60$  $\bm{k}$-mesh, respectively.
The calculated Fermi surface was depicted using FermiSurfer\cite{Kawamura2019Comp}.  
%To estimate the Fermi energy from the experimentally determined cross section of the Fermi surface, we employed a $250\times 250$ $\bm{k}$-mesh on the $k_z=0$ plane using our tight-binding model.

\section{RESULTS and DISCUSSIONS}

%Figure1==========================================================================

%\onecolumngrid

\begin{figure}[t]
	\begin{center}
		\includegraphics[width=1\linewidth]{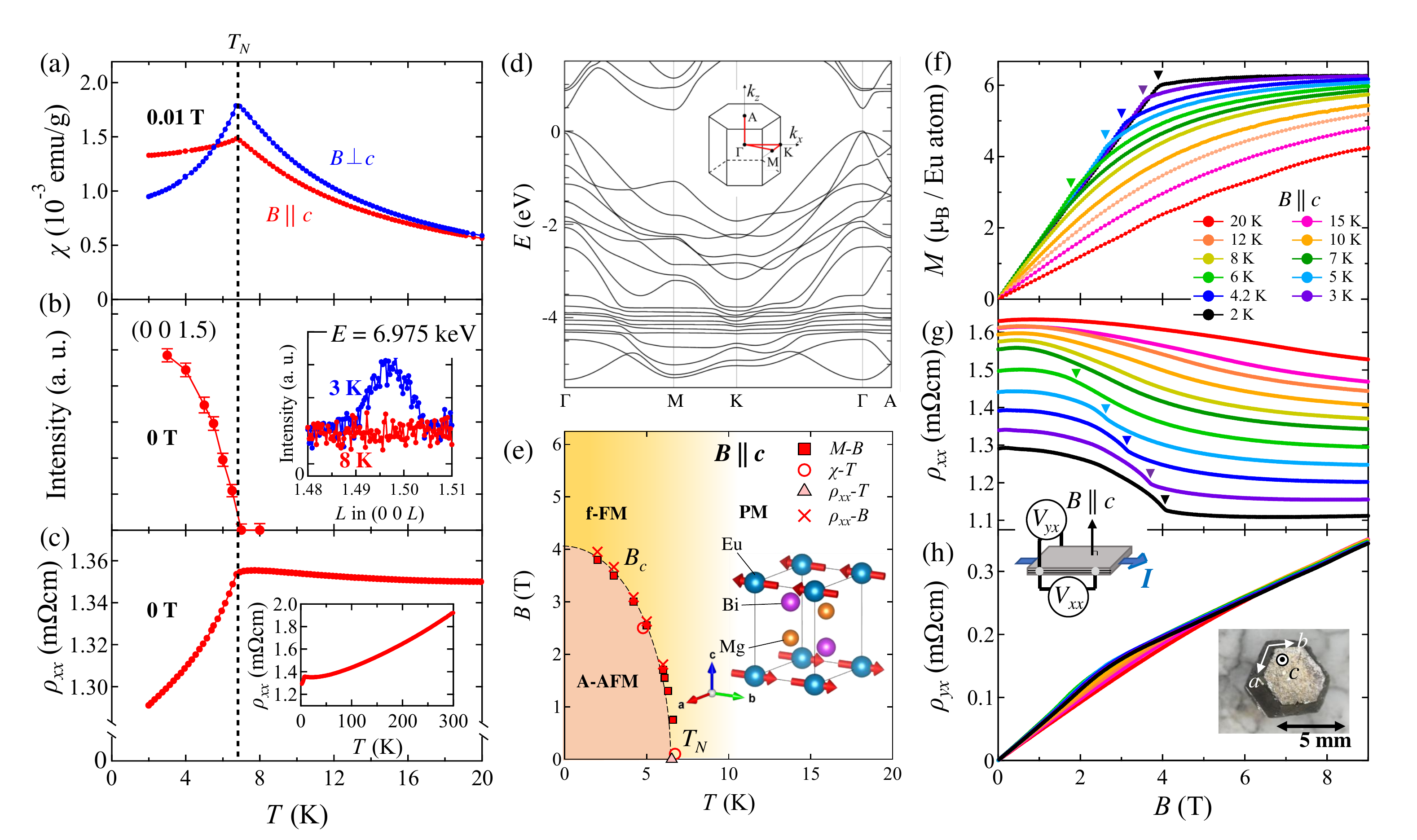}
		\caption[Structure]{\label{fig:phase}
%		The inset shows a photograph of a typical EuMg$_2$Bi$_2$ single crystal. 
%		The large hexagonal-like facet corresponds to the $ab-$plane. 
		(a) Temperature dependence of out-of-plane (red) and in-plane (blue) magnetic susceptibilities measured at 0.01 T in a cooling run. 
		The vertical dotted line denotes the Neel temperature $T_N$($=6.7$ K) for Eu sublattice. 
		(b)Temperature dependence of intensity of resonant magnetic reflection (0 0 1.5) at $E =$ 6.975 keV at 0 T. 
		The inset shows the intensities of (0 0 1.5) magnetic reflection below and above $T_N$ (3 K and 8 K, respectively). 
		(c) The temperature dependence of in-plane resistivity $\rho_{xx}$ below 20 K. 
		The inset shows overall temperature dependence of $\rho_{xx}$.
		%The $\rho_{xx}$ shows metallic behavior below 300 K as shown in inset of (c). 
		(d) Calculated band structure of EuMg$_2$Bi$_2$ for the A-type antiferromagnetic order of Eu spins.
		The inset shows the 1st Brillouin zone.		
		%(a) The crystal and magnetic structure of EuMg$_2$Bi$_2$.  
		(e) The magnetic phase diagram of EuMg$_2$Bi$_2$ as functions of the field ($B\parallel c$) and temperature ($T$). PM, A-AFM and f-FM denote the paramagnetic, A-type antiferromagnetic and forced-ferromagnetic phases, respectively. 
		$B_c$ corresponds to the transition field to the f-FM phase. 
		The inset exhibits the crystal and magnetic structure of EuMg$_2$Bi$_2$.
		The easy-axis direction is here assumed to be parallel to the $b$-axis because the in-plane spin direction cannot be determined from the present experiment on a sample with magnetic domains\cite{EMgB_neutron} (Fig. S2 for details)
%		The direction of Eu spins within the $ab$-plane is not determined experimentally.
		(f)-(h) The field dependence of (f) magnetization $M$, (g) $\rho_{xx}$ and (h) Hall resistivity $\rho_{yx}$ (up to 9 T).
		 Triangles in (f) and (g) denote $B_c$ at various temperatures below $T_N$.
		 All $\rho_{xx}$ data in (g) are shifted vertically by 20 or 40 $\mathrm{\mu\Omega}$cm for clarity. 
		 The insets show a schematic illustration of measurements of $\rho_{xx}$($V_{xx}$) and $\rho_{yx}$($V_{yx}$) and a photograph of a typical crystal.}
	\end{center}
\end{figure}

%\twocolumngrid

%%%%%%%%%%%%%%%%%%%%%%%%%%%%%%%%%%%%%%%%%%%%%%%%%%%%%%%%%%%%%%%%%%%%%%%%%%%%%%%%%%%%%

Figure 1(a) shows temperature dependence of the magnetic susceptibility, where a cusp-like anomaly is discernible at $T_N=6.7$ K, indicating an antiferromagnetic transition of the Eu sublattice.
While the magnetic susceptibility in $B\parallel c$ is almost constant below $T_N$, that for $B\perp c$ decreases significantly below $T_N$.
This suggests that the easy axis of magnetization of Eu spins is within the $ab$-plane.
To investigate the order structure of Eu spins in the ground state, we performed the resonant X-ray magnetic scattering measurement.
As shown in inset of Fig. 1(b), a superlattice reflection (0 0 1.5) is observed below $T_N$, resonating at the Eu $L_{3}$ absorption edge $E=6.975$ keV\cite{Masuda2016SA,Masuda2020PRB} [Fig. S2(a) for details].
Since the intensity of this peak develops significantly below $T_N$ [Fig.1(b)], it arises from the magnetic reflection due to the Eu spin order.
Furthermore, $q$-vector of (0 0 0.5) indicates that the Eu spins show A-type antiferromagnetic (A-AFM) order with cell doubling along the $c$-axis direction.
This result is consistent with previous neutron diffraction experiments at zero field\cite{EMgB_neutron}.
By considering the anisotropy of magnetic susceptibility, the order structure of Eu spins is determined as shown in inset of Fig. 1(e).
%It should be noted that 
%
%
%

%%%%%
\par
%%%%%

Figure 1(d) shows the band structure of EuMg$_2$Bi$_2$ in the A-AFM phase calculated using the HSE06 functional, which shows a direct band gap $E_g\sim0.5$ eV at the $\Gamma$ point.
This semiconducting band structure is similar to that calculated for $X$Mg$_2$Bi$_2$ ($X=$Sr, Ba)\cite{XMB_calc_HSE06, BaMg2Bi2_ARPES_Sato}.
However, the metallic conduction was experimentally observed from room temperature to 20 K [inset to Fig. 1(c)], and thereby, indicating that a few carriers are doped in reality.
As shown in Fig. 1(c), the in-plane resistivity $\rho_{xx}$ slowly increases with decreasing temperature below 20 K, followed by a rapid decrease below $T_N=6.7$ K.
Thus, doped carriers are strongly coupled with the antiferromagnetic order of the Eu spins.

%%%%%
\par
%%%%%

To investigate the coupling between the magnetism and transport properties, we applied the field along the $c$-axis to measure the magnetization and resistivity.
Figure 1(f) shows field dependence of magnetization $M$ below 20 K.
At 2 K, $M$ increases almost linearly with increasing field up to $B_c=3.9 $ T, above which $M$ is almost constant, and Eu spins are fully polarized along the $c$- axis.
% i.e., forced ferromagnetic (f-FM) phase.
The saturated magnetization in this forced ferromagnetic (f-FM) phase is $M_{\mathrm{sat}}\sim6.1\mu_{\mathrm{B}}$/Eu, which is close to the magnetic moment  of bare Eu$^{2+}$ ion.
This indicates that the Eu $4f$ electrons are almost localized.
As temperature increases, $B_c$ shifts to lower magnetic fields and disappears above $T_N$.
The resultant magnetic phase diagram is shown in Fig. 1(e).

%%%%%
\par
%%%%%

Figure 1(g) shows the field dependence of $\rho_{xx}$ below 20 K for $B\parallel c$.
At 2 K, the negative magnetoresistance was observed for $B<B_c$.
A clear kink is discernible at $B_c$, above which $\rho_{xx}$ is almost constant up to 9 T.
As temperature increases, $B_c$ shifts to lower magnetic fields and the kink becomes less apparent.
Figure 1(h) shows the magnetic field dependence of the Hall resistivity $\rho_{yx}$ for $B\parallel c$.
The positive slope of $\rho_{yx}$ indicates that hole carriers are slightly doped, and thus, $E_F$ crosses the top of the valence band at $\Gamma$ point.
Although $\rho_{yx}$ is almost linear with respect to field above $T_N$, it deviates from the straight line below $T_N$; a hump structure at approximately 2.5 T becomes pronounced as temperature decreases.
This signals the evolution of AHE coupled with Eu magnetic ordering.

%Figure4=AHE=========================================================================

\begin{figure}[t]
	\begin{center}
		\includegraphics[width=0.8\linewidth]{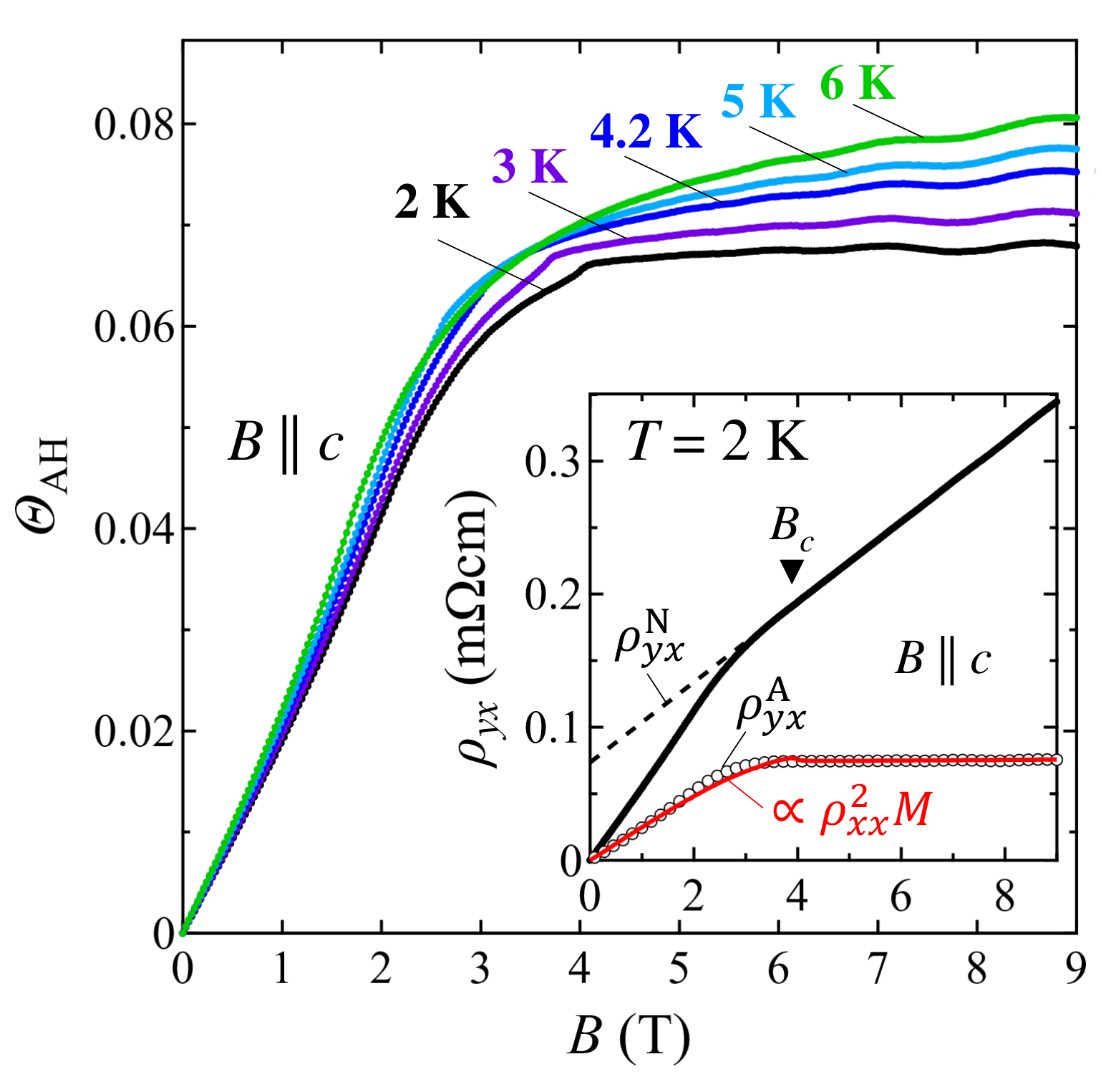}
		\caption[Structure]{\label{fig:AHE}
		Field dependence of anomalous Hall angle $\Theta_{\mathrm{AH}} = \rho_{yx}^{\mathrm{AH}}/\rho_{xx}$ at various temperatures. 
		The oscillatory structure above 6 T is due to SdH oscillation for $\rho_{yx}$. 
		The inset shows the field dependence of $\rho_{yx}$ and its anomalous part $\rho_{yx}^{\mathrm{A}}$ at 2 K. 
		The black dotted line corresponds to the ordinary part $\rho_{yx}^{\mathrm{N}}$ of $\rho_{yx}$ at 2 K, which is obtained from the liner fit to the experimental data above $B_c$. 
		The red curve represents the fitted result for $\rho_{yx}^{\mathrm{A}}$.
		}
	\end{center}
\end{figure}

%%%%%%%%%%%%%%%%%%%%%%%%%%%%%%%%%%%%%%%%%%%%%%%%%%%%%%%%%%%%%%%%%%%%%%%%%%%%%%%%%

To extract the anomalous component $\rho_{yx}^{\mathrm{A}}$, we estimate the ordinary component $\rho_{yx}^{\mathrm{N}}=R_{\mathrm{H}}B$ from the data above $B_c$, where the magnetization is fully saturated [Fig. 1(f)].
As shown in the inset to Fig. 2, the Hall coefficient $R_{\mathrm{H}}$ is unambiguously determined at 2 K from the slope of $\rho_{yx}$ above $B_c$.
The resultant $\rho_{yx}^{\mathrm{A}}=\rho_{yx}-\rho_{yx}^{\mathrm{N}}$ is appropriately fitted by $\rho_{yx}^{\mathrm{A}}\propto \rho_{xx}^2M$ (inset to Fig. 2).
This suggests that the intrinsic contribution from the Berry curvature is dominant in the AHE for EuMg$_2$Bi$_2$\cite{MnGe_AHE, AHE_review}.
Figure 2 shows the field dependence of anomalous Hall angle $\Theta_{\mathrm{AH}}=\rho_{yx}^{\mathrm{A}}/\rho_{xx}$ below $T_N$ (see Fig. S3 for determining $\rho_{yx}^{\mathrm{A}}$ at each temperature).
At $B>B_c$, the $\Theta_{\mathrm{AH}}$ is nearly constant ($\sim0.068-0.080$), reflecting the field dependence of magnetization.
The constant value of $\Theta_{\mathrm{AH}}$ is comparable to those reported for ferromagnetic Weyl semimetals Co$_3$Sn$_2$S$_2$ ($\sim0.2$ at 150 K)\cite{Co3Sn2S2_AHE_Felser}, Fe$_3$GeTe$_2$ ($\sim0.09$ at 2 K)\cite{Fe3GeTe2_AHE_Kim} and GdPtBi ($\sim$0.15 at 2.5 K)\cite{GPB_Suzuki}.
Hence, such a large $\Theta_{\mathrm{AH}}$ implies that EuMg$_2$Bi$_2$ has band crossing points near $E_F$ in the f-FM phase as a source of Berry curvature.
By taking advantage of the simple band structure with a single valley at the $\Gamma$ point [Fig. 1(d)], we reveal its relation to the observed large AHE below.
%
%

%Figure3=Band calculation=========================================================================

%\onecolumngrid

\begin{figure}[t]
	\begin{center}
		\includegraphics[width=1\linewidth]{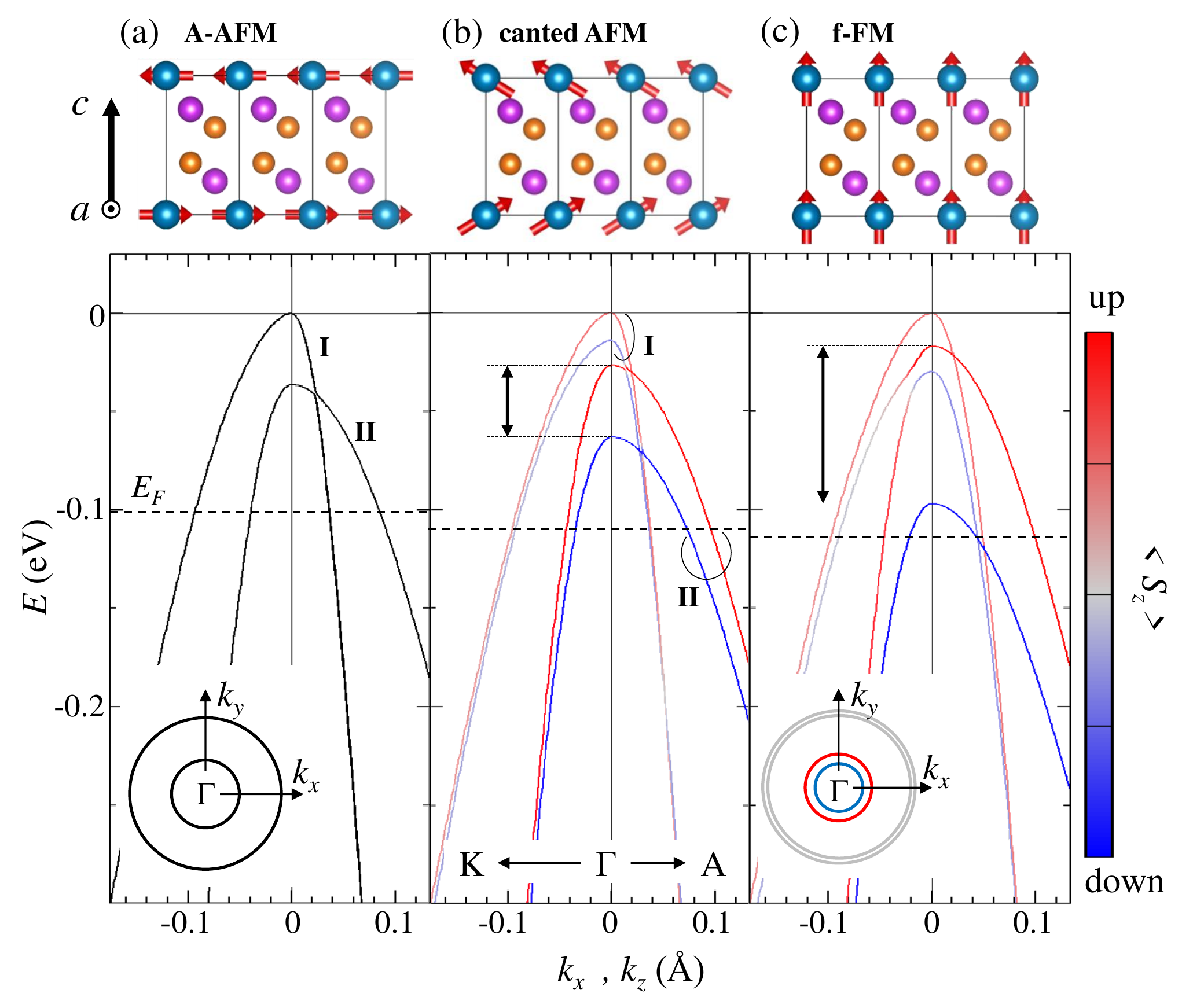}
		\caption[Structure]{\label{fig:band}
		(a)-(c)Valence band structures around the $\mathrm{\Gamma}$ point near $E_F$ (lower panels) for various magnetic states (upper panels).
		Red and blue colors of band dispersions represent spin up and down, respectively. 
		The dotted lines denote the Fermi energy $E_F$ determined from the SdH frequency $B_F$ at $T =$ 1.4 K. 
		Insets of (a) and (c) show the schematic illustration of Fermi surfaces within the $k_xk_y$-plane, where the color denotes the $\langle S_z\rangle$ value.
		}
	\end{center}
\end{figure}

%\twocolumngrid

%%%%%%%%%%%%%%%%%%%%%%%%%%%%%%%%%%%%%%%%%%%%%%%%%%%%%%%%%%%%%%%%%%%%%%%%%%%%%%%%%

To clarify the impact of Eu spins on the band structure, we performed the first-principles band calculation for various magnetic states.
We show the variation of valence bands near the $\Gamma$ point in Figs. 3(a)-(c).
In the A-AFM phase (at zero field), there are two valence bands I and II consisting of the Mg-$s$ and Bi-$p$ orbitals, which cross each other at $k_z\sim0.03$\AA$^{-1}$ on the $\Gamma$-A line [Fig.\,3(a)].
This leads to a Dirac-like point because the crossing bands are spin-degenerate in the A-AFM phase with no net magnetization.
When the field is applied along the $c$-axis, Eu spins start to cant towards the $c$-axis.
In Fig.\,3(b), we show the band structure calculated for the Eu spins tilted by approximately 30$^{\circ}$ with respect to the $ab$-plane [upper panel of Fig. \ref{fig:band}(b)].
In this canted AFM phase, the net magnetization from the Eu spins is no longer zero, which leads to clear spin splitting (red and blue) in bands I and II owing to the exchange interaction\cite{KSTakahashi2019SA,EuP3_Mayo,EuMnBi2_PRB}.
This leads to multiple Weyl-like points, where the spin-polarized bands cross.
The magnitude of splitting increases as the field or the net magnetization along the $c$-axis increases.
In the f-FM phase [Fig. 3(c)], the splitting of the band II reaches as large as $\sim$80 meV, where the Weyl-like points are formed at $E\sim-20$ meV and $-120$ meV.

%Figure2=Quantum oscillation=========================================================================
\begin{figure}[t]
	\begin{center}
		\includegraphics[width=1\linewidth]{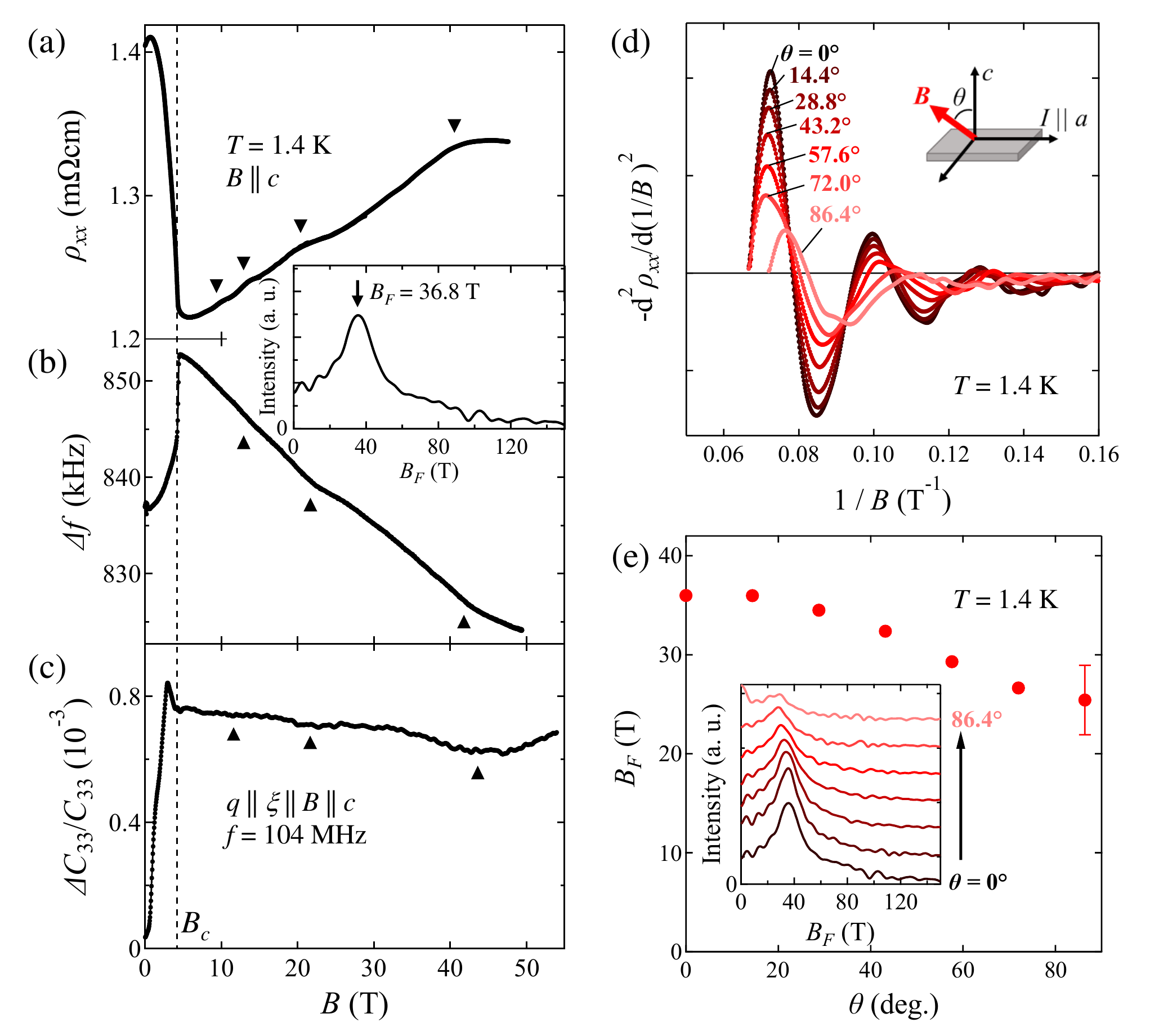}
		\caption[Structure]{\label{fig:QO}
		(a)-(c) Field dependence of (a) $\rho_{xx}$, (b) resonance frequency of tunnel diode oscillator $\Delta f$ and (c) the relative change of longitudinal elastic constant $\Delta C_{33}/C_{33}$ at $T =$ 1.4 K up to $\sim$50 T. 
		Triangles denote the oscillatory component. 
		The inset shows the fast Fourier-transform (FFT) spectrum of $-d^2\rho_{xx}/d(1/B)^2$ for the field of 5-15 T. 
		$q$ and $\xi$ in (c) indicate the propagation and polarization direction of ultrasonic waves, respectively.
		(d) $-d^2\rho_{xx}/d(1/B)^2$ versus $1/B$ at various field tilt angles ($\theta$), where $\theta$ is the angle between the $c-$axis and magnetic field (see inset). 
		(e) The $\theta$ dependence of oscillation frequency $B_F$, which is determined from the FFT spectra of $-d^2\rho_{xx}/d(1/B)^2$ shown in the inset. }
	\end{center}
\end{figure}
%%%%%%%%%%%%%%%%%%%%%%%%%%%%%%%%%%%%%%%%%%%%%%%%%%%%%%%%%%%%%%%%%%%%%%%%%%%%%%%%%

%%%%%
\par
%%%%%

To discuss their actual effect on the AHE, it is necessary to determine the position of $E_F$.
Hence, we performed high-field measurements up to 55 T using a pulsed magnet to reveal the quantum oscillation phenomena.
Figure 4(a) shows the field dependence of $\rho_{xx}$ at $T = 1.4$ K.
At fields exceeding 10 T, $\rho_{xx}$ increases almost linearly with respect to the field wherein the SdH oscillation is superimposed (solid triangles indicate the peak positions of the oscillation).
The fast Fourier-transform (FFT) spectrum of $-d^2\rho_{xx}/d(1/B)^2$ shows the clear single peak at the frequency of $B_F=36.8$ T as shown in the inset.
Figures 4(b) and (c) show the field dependence of resonance frequency of TDO, $\Delta f$, and the relative elastic constant, $\Delta C_{33}/C_{33}=[C_{33}(B)-C_{33}(B=0)]/C_{33}(B=0)$, respectively.
For both physical quantities, the oscillatory structures consistent with the SdH oscillation were observed above $B_c$ (denoted by solid triangles).
Moreover, by considering that the acoustic de Haas-van Alphen effect in Fig.\,4(c) is due to the Landau quantization via the electron-phonon coupling\cite{CeRhIn5_PRB_AdHvA, Kataoka_AdHvA}, the observed quantum oscillation should arise from a bulk band.

%%%%%
\par
%%%%%

The quantum oscillation of bulk origin is also supported by the dependence of $B_F$ on the tilt angle of field.
Figure 4(d) shows $-d^2\rho_{xx}/d(1/B)^2$ versus $1/B$ at $T = 1.4$ K for various $\theta$ values, where $\theta$ denotes the tilt angle with respect to the $c$-axis (see the inset).
While the amplitude of SdH oscillation gradually decreases as $\theta$ increases, the positions of peaks and dips are almost unchanged with respect to $\theta$, resulting in nearly $\theta$-independent $B_F$ [Fig.\,4(e)].
This $\theta$ dependence of $B_F$ indicates that the detected Fermi surface is 3D and almost isotropic\cite{QO_Shoenberg}, and it is consistent with the hole pocket at the $\Gamma$ point obtained by the first-principles calculation (Fig. S4).

%%%%%
\par
%%%%%

Finally, we estimate the position of $E_F$ based on the quantum oscillation.
Since the quantum oscillation is discernible in the f-FM phase (above 10 T), we compare the experimental result with the band structure for the f-FM phase [Fig. 3(c)].
From the experimental $B_F$ value, the Fermi surface cross section $S_F$ on the $k_xk_y$-plane is estimated as $S_F=0.352$ nm$^{-2}$ at $\theta=0^{\circ}$ [by Onsager's theorem, $S_F=(2\pi e/\hbar)B_F$].
Theoretically, the Fermi surface on the $k_xk_y$-plane approximately consists of the inner and outer surfaces, ignoring the spin splitting [inset to Fig. 3(c)].
Assuming that the observed $B_F$ corresponds to the inner Fermi surface, $E_F$ is located around $E=-110$ meV\cite{footnote_calc}, and the carrier density is calculated as $n_{\mathrm{calc}}\sim1.1\times10^{19}$ cm$^{-3}$.
This carrier density is in good agreement with that estimated from the Hall coefficient $R_{\mathrm{H}}$ at high magnetic field ($n_{\mathrm{Hall}}=2.1\times10^{19}$ cm$^{-3}$ at 2 K).
Conversely, if we assume that the experimental $B_F$ corresponds to the outer surface, then we can obtain $E_F=-40$ meV and $n_{\mathrm{calc}}\sim1.1\times10^{18}$ cm$^{-3}$.
The latter value is 1/10 lower than $n_{\mathrm{Hall}}$.
Therefore, the quantum oscillation should arise from the inner Fermi surface, and the resultant $E_F$ ($=-$110 meV) is denoted by the horizontal dashed line in Fig. 3(c).

%%%%%
\par
%%%%%

Based on the fact that the carrier density is constant irrespective of the magnetic order, $E_F$ was also determined for the A-AFM and canted AFM phases [dashed lines in Fig.3(a) and Fig.3(b)].
In the A-AFM phase, $E_F$ is located approximately 60 meV below the Dirac point.
Conversely, in the f-FM phase, the lower-energy Weyl point, which is generated from the Dirac point by spin splitting, is located very close to $E_F$.
In such a state, the conduction carriers can be significantly affected by the Berry curvature even at low temperatures, leading to the large AHE shown in Fig. 2.
To test this mechanism, it is important to examine the dependence of AHE on the carrier density in future studies.

%%%%%
\par
%%%%%

%%Table1==============================================================================
%%
%%\begin{table}[h]
%%	\begin{center}
%%		\caption[Structure]{The anomalous Hall angle for various magnet.}
%%		\begin{tabular}{*{3}{c@{\hskip2zw}}c}
%%			Material&$\Theta_{\mathrm{AH}}$&Reference \\
%%			\hline\hline
%%			EuMg$_2$Bi$_2$ & 0.068 & This work\\
%%			Co$_3$Sn$_2$S$_2$ & 0.035 & \cite{Co3Sn2S2_AHE}\\
%%			Fe$_3$GeTe$_2$ & 0.042-0.086 & \cite{Fe3GeTe2_AHE}\\
%%			MnSi & 0.01 & \cite{MnSi_AHE}\\
%%			MnGe & 0.012 & \cite{MnGe_AHE}\\
%%			Fe & 0.011 & \cite{Fe_AHE}\\
%%		\end{tabular}
%%	\end{center}
%%\end{table} 
%%
%%%%%%%%%%%%%%%%%%%%%%%%%%%%%%%%%%%%%%%%%%%%%%%%%%%%%%%%%%%%%%%%%%%%%%%%%%%%%%%%%%%%%%
%%

In conclusion, we examined the band structure controllable by the magnetic order and associated large anomalous Hall effect (with the Hall angle of 0.068-0.080) in degenerate magnetic semiconductor EuMg$_2$Bi$_2$, which exhibits the same crystal structure as putative topological material (Sr, Ba)Mg$_2$Bi$_2$.
At zero field below $T_N=6.7$ K, EuMg$_2$Bi$_2$ exhibits the A-type antiferromagnetic order with Eu spins along the $ab$-plane.
In this antiferromagnetic phase, the first principles calculation predicted a simple semiconducting band structure with a single valley at the $\Gamma$ point.
However, when field is applied along the $c$-axis, the bands exhibit large spin splitting due to the exchange interactions with Eu spins.
Consequently, the number and position of band crossing points (Weyl points) vary significantly depending on the inclination angle of the Eu spins with respect to the $ab$-plane.
Based on the experimental $E_F$ determined from the quantum oscillation at high fields, we revealed that one of the Weyl points is formed very close to $E_F$ in forced ferromagnetic phase, and this can likely lead to the large anomalous Hall effect in this compound.

%\section*{$*$Corresponding author}
%$^*$kondo@gmr.phys.sci.osaka-u.ac.jp, $^\dagger$sakai@phys.sci.osaka-u.ac.jp
%
%\section*{$*$Acknowledgements}
\begin{acknowledgments}
The authors are grateful to Y. Kohama for the experimental advices.
This work was partly supported by the JSPS KAKENHI (Grant Nos. 19H01851, 19K21851, 22H00109, 22J10851, 21H00147) and the Asahi Glass Foundation.
The X-ray diffraction experiments at KEK were performed under the approval of the Photon Factory Program Advisory Committee (Proposal Nos. 2018S2-006 and 2021S2-004).
This work was carried out, in part, at the Center for Spintronics Research Network (CSRN), Graduate School of Engineering Science, Osaka University.
\end{acknowledgments}

\section*{Data availability}
The data that support the findings of this study are available from the corresponding author upon reasonable request.


\begin{thebibliography}{99}
\bibitem{footnote_calc} We determined the Fermi energy ($E_F$) in Figs. 3(b) and 3(c) so that the experimental Fermi-surface cross section is consistent with the averaged cross section of the two valence bands labeled as II.
\end{thebibliography}
\end{document}